\documentclass[10pt,showpacs,preprintnumbers,amsmath,amssymb]{revtex4}
\usepackage{units}
\usepackage{bbold}
\usepackage{graphicx}
\usepackage{dcolumn}
\usepackage{bm}
\usepackage{bbm}
\usepackage{slashed}

\begin{document}

\title{Quaternionic particle in a relativistic box}
\vspace{2cm}
\author{Sergio Giardino}
\email{p12@ubi.pt}
\affiliation{ Departamento de F\'isica  \& Centro de Matem\'atica e Aplica\c c\~oes\\ Universidade da Beira Interior,
Rua Marqu\^es D'\'Avila e Bolama 6200-001 Covilh\~a, Portugal}

\begin{abstract}
\noindent This study examines Quaternion Dirac solutions for an infinite square well. The
quaternion result does not recover the complex result within a particular limit. This raises the possibility that quaternionic
quantum mechanics may not be understood as a correction to complex quantum mechanics, but it may also be a structure that can be used 
to study phenomena that cannot be described through the framework of complex quantum mechanics.
\end{abstract}

\maketitle

\section{Introduction}

A physical theory may be better understood if it is generalized, for several reasons. 
Firstly, if a theory is understood as
a particular case of a more general theory, then its limits are well established, and the variety of phenomena that may 
be described by this particular theory are also well defined. On the other hand, the general theory may have several theories at 
particular limits, and relations between particular cases of the general theory may be established. An example of this occurs in
string theory, where different theories are connected by duality transformations (T-dualities).

However, this understanding may not easily be reached if the generalized theory introduces additional constraints. In this case,
a solution of the generalized theory may not recover a well-known solution of the particular solution within a specific limit. As we
know, the quaternion Dirac equation reduces to the complex Dirac equation if the quaternionic part of the potential goes to
zero \cite{DeLeo:2013xfa}. However, the quaternion case may have more constraints from boundary conditions. This may not be the case for the 
scattering of Dirac particles by a step potential \cite{DeLeo:2015qsp}, and the complex solution is recovered from the quaternion solution. 
On the other hand, as we shall see, the quaternion Dirac solution for the infinite square well does not recover the complex Dirac 
solution for the infinite square well. This means that the interpretation of the physical problem in the complex situation may not be 
analogous to the interpretation of the physical problem in the quaternionic situation.

In order to introduce our problem, we remember that a quaternion generalization of quantum mechanics is obtained by allowing the
 wave-function, which is evaluated over the
complex number field, to be evaluated in the quaternion number field. Writing a quaternionic wave-function in the symplectic notation, 
we get
\begin{equation}\label{psi}
\Psi=U+jW,
\end{equation}
so that $U$ and $W$ are functions evaluated over complex numbers and $j$ is one of the three quaternionic anti-commuting imaginary 
unities. The third  imaginary unit is defined by $k=ij$, so that a general quaternion $Q$ is written with real coefficients such as 
$Q=A+Bi+Cj+Dk$. 
A quaternionic wave-function has more degrees of freedom than a complex wave-function, and it may have some probability
density coming from the pure quaternionic term $jW$ of the wave-function (\ref{psi}). 

Quaternionic generalization of 
quantum mechanics and of quantum field theory has been studied and collected by Adler \cite{Adler:1995qqm}. 
However, Adler's work is focused on formal aspects of quaternionic quantum mechanics, and specific examples are still required
in order to determine whether there are measurable quaternionic effects. In the
case of relatisvitic quantum mechanics, Davies \cite{Davies:1990pm} studied Lorentz invariant scalar potentials,
and showed that this kind of potential may generate representation-dependent solutions. Another interesting result 
from Davies is the decoupling of the dynamics of the quaternionic part of the wave-function from the dynamics of the 
complex part of the wave-function. This raises the possibility that quaternionic effects may be impossible to be observed,
and hence physically irrelevant.
On the other hand, for a constant vector potential \cite{DeLeo:2013xfa}, the complex part is not decoupled from the
quaternionic part, and then quaternionic effects may, in principle, be detected. This possibility may be observed based on several
features of the solution. First of all, it has been 
ascertained that, in accordance with complex Dirac wave-functions, quaternionic Dirac wave-functions also 
have three energy regimes named the diffusion zone, the evanescent zone and the Klein zone. A measurable quaternion effect 
that may be observed is the energy range of the evanescent zone, which is sensitive to the value of the pure quaternionic part 
of the potential. Another interesting feature of this solution concerns the effect of the pure quaternionic term of the potential
on the speed of the generated particles and anti-particles. The pure imaginary complex
part of the potential has the effect of decelerating particles and accelerating anti-particles. In other words, we say that the usual
barrier has an repulsive effect on particles and an attractive effect on anti-particles. We can suppose that this attractive character of 
the potential is responsible for generating Dirac anti-particles in the Klein zone. The pure imaginary  quaternion
component of the potential has an opposite effect, accelerating anti-particles and decelerating particles, and generating 
particles for energies belonging to the range of the Klein zone.

In this article we apply the solutions of the quaternion Dirac equation with a vector potential \cite{DeLeo:2013xfa} to
the infinite one-dimensional square well, a system that confirms that the quaternionic part of the potential generates 
measurable effects on wave-function. The relativistic infinite square well has already been studied for the complex case in various dimensions
\cite{Fiolhais:1998diw,Alberto:2011wn} and also for a finite square \cite{Alhaidari:2009ug}. We extend it to the one-dimensional
quaternion case. It must be said that many confining solutions for the complex Dirac equation have been developed, and some 
are already part of a textbook \cite{Greiner:1990tz}, and there are a variety of studies on different aspects of this
subject, such as the boundary conditions \cite{DeVincenzo:1997bcb,DeVincenzo:1999sbb,Belyi:2004dpb} and
confining solutions in scalar potentials \cite{Nogami:1988bsd,Nogami:2003dpb}. None of these cases have been generalized for the
quaternionic situation, accordingly this article is just a first step in understanding quantum relativistic confined solutions.

In Section \ref{S2} we describe the quaternion Dirac equation and its solutions given from \cite{DeLeo:2013xfa}. In Section \ref{S3} 
we describe the model and present our results. Section \ref{S4} presents our conclusions and future perspectives.

\section{ The quaternionic Dirac solution for a step potential\label{S2}}
The anti-hermitian Dirac equation for an arbitrary quaternionic potential in the natural system of units \cite{DeLeo:2013xfa} is
\begin{equation}\label{dir_eq}
\partial_t\,\Psi(\bm
r,\,t)=-\Big[\bm{\alpha\cdot\nabla}+im\,\beta+\bm{h\cdot V}(\bm r)\mathbbm{1} \Big]\,\Psi(\bm r,\,t).
\end{equation}
$V_n(\bm r)$ are the real components of $\bm V$ for $n=\{1,2,3\}$  and
the complex units are contained in $\bm h=(i, j, k)$. The $4\times 4$ matrices
$\mathbb{1}$, $\alpha_m$ and $\beta$ satisfy the algebra
\begin{equation}
\alpha_m=\alpha_m^\dagger,\qquad\beta=\beta^\dagger,\qquad\alpha^2=\beta^2=\mathbbm{1}
,\qquad\{\beta,\,\alpha_m\}=0\qquad\mbox{and}\qquad\{\alpha_m,\,\alpha_n\}=2\delta_{mn}\,\mathbbm{1},
\end{equation}
where $\mathbb{1}$ is the $4\times 4$ unity matrix. The representation adopted for the calculations is
\begin{equation}
\alpha_m=\left(
\begin{array}{cc}
0&\sigma_m\\
\sigma_m&0\\
\end{array}\right),\qquad
\beta=\left(
\begin{array}{cc}
\mathbbm{1}_2&0\\ 
0&-\mathbbm{1}_2\\
\end{array}
\right),
\end{equation}
where $\mathbbm{1}_2$ represents a $2\times 2$ unity matrix and $\sigma_m$ represents the $2\times 2$ Pauli matrices for $m=\{1,\,2,\,3\}$. The most general potential for 
the Dirac equation may be written as \cite{deCastro:2003zz}
\begin{equation}\label{pot_gen}
V=A_t(x)\mathbb{1}+A_i(x)\alpha+ U(x)\beta + \Phi(x) \beta \gamma^5,
\end{equation}
where $A_t$ is and $A_i$ are components of the four-potential and $U$ and $\Phi$ are real-valued functions. $U$ generates the
scalar term of the potential and $\Phi$ generates the pseudo-scalar term. We consider a step potential
\begin{equation}\label{pot}
\bm V(\bm r,\,t)=\left\{\;
\bm V\mbox{ for } z>0\; \mbox{ and }\; 0\mbox{ for } z<0\;\right\}.
\end{equation}
with $\bm V=(V_1,\,V_2,\,V_3)$ a real constant vector. This potential can be identified with $A_t$ of (\ref{pot_gen}), and then 
represents a vector potential where the only non-zero component is the time component. The pure quaternionic term of the potential
may be rewritten as $V_1=V_0$ and $W_0=V_3+iV_2$, where, of course,
\begin{equation}\label{phase}
W_0=|W_0|e^{i\phi}, \qquad\mbox{with}\qquad 
|W_0|=\sqrt{V_2^2+V_3^2} \qquad\mbox{and}\qquad
\phi=\arctan\frac{V_2}{V_3}.
\end{equation}

The system is then composed of particles that move one-dimensionally and whose solution of the Dirac equation \cite{DeLeo:2013xfa}
is  written in general as
\begin{equation}
\Psi_\pm=(U+jW)e^{i(Q_\pm z-Et)}
\end{equation}
so that
\begin{eqnarray}
&& Q_\pm^2=q_\pm^2+|W_0|^2\pm 2\delta.\label{momentum},\qquad
\mbox{where}\\
&& q_\pm^2=\big(E\pm V_0\big)^2-m^2,\qquad\delta=\sqrt{E^2V_0^2+p^2|W_0|^2}-E\,V_0,\qquad\mbox{and}\qquad\ p^2=E^2-m^2.
\end{eqnarray}
The  solution whose momentum is $Q_-$ is similar to the complex solution of the Dirac equation, where there are three
characteristic energy zones, namely the diffusion zone, the evanescent zone and the Klein zone. One of the differences between the 
cases is that the range of the evanescent zone for the quaternion case is given by
\begin{equation}\label{zeta_tunel}
\Delta E=\sqrt{|W_0|^2+(m+V_0)^2}-\mbox{max}\big[m,\,\sqrt{|W_0|^2+(m-V_0)^2}\big].
\end{equation}
The energy range of the evanescent zone depends on the quaternionic term of the potential. On the other hand, the
solution whose momentum is given by $Q_+$ has only the diffusion zone, and  this energy zone is inhabited only by anti-particles. This is another difference
in relation to the complex case, because the complex case presents a coexistence between particles and anti-particles with the same energy and different
momenta. Finally, the wave-functions for the one-dimensional step are
\begin{equation}\label{psi_step}
\Psi_-=\left[
\begin{array}{c}
\big(1-j\,W_0\,M_-\big)\chi\\\big(A_--j\,W_0\,N_-\big)\sigma_3\chi
\end{array}
\right]e^{i(Q_{-}z-Et)}\qquad\mbox{and}\qquad
\Psi_+=\left[
\begin{array}{c}
\big(A_+-j\,\overline W_0\,N_+\big)\sigma_3\chi\\\big(1-j\,\overline{ W}_0\,M_+\big)\chi
\end{array}
\right]e^{i(Q_{+}z-Et)}.
\end{equation}
$\chi=\{(1,\,0)^t,\,(0,\,1)^t\}$ are two-dimensional spinors, $\overline W_0$ is the complex conjugate of $W_0$ and  
\begin{equation}\label{AMN}
A_\pm=\frac{Q_\pm}{E\pm V_0+m\pm\frac{\delta}{E-m}},\qquad
M_\pm=\frac{E\mp V_0-m+Q_\pm A_\pm}{q_\mp^2-Q_\pm^2}\qquad\mbox{and}\qquad
N_\pm=\frac{Q_\pm+A_\pm\big(E\mp V_0+m\big)}{q_\mp^2-Q_\pm^2}.
\end{equation}
Now we use the step potential solution to build a confined solution.

\section{\label{S3} The quaternionic confined Dirac solutions}

The step solution presented in the former section has two possible momenta, and consequently we must consider two confined 
solutions, one for each momentum. However, in order to obtain a clearer 
understanding of the situation, we consider the non-relativistic limit which gives the usual quantum infinite square well

\subsection{The non-relativistic limit}
We consider the potential (\ref{pot}) as given by $\bm V=(V_0,\,V_2,\,V_3)$, so that
\begin{equation}\label{NR_pot}
 V_0(\bm r,\,t)=\left\{\;
 0\mbox{ for } 0<z<L\; \mbox{ and }\; \infty\mbox{ otherwise }\right\},
\end{equation}
where $L$ is a real number that gives the width of the well. There is a constant and finite pure quaternionic potential inside
 the well, and in this situation the parameters of the solutions are
\begin{equation}\label{NR_param}
Q_\pm=p\pm|W_0|,\qquad A_\pm=\frac{p}{E+m},\qquad M_\pm=\mp\frac{1}{|W_0|}\frac{p}{E+m}\qquad\mbox{and}\qquad N_\pm=\mp\frac{1}{|W_0|}.
\end{equation}
We use $p>|W_0|$. For $|W_0|<p$ some signs must be flipped, but there is no change in the physical interpretation. In the non-relativistic
limit, $Q_\pm\ll m$, and then $A_\pm,\,M_\pm\to 0$. The wave-functions (\ref{psi_step}) in this non-relativistic regime are
\begin{equation}\label{Psi_pm}
\Psi_+=\left(
\begin{array}{c}
je^{-i\phi}\sigma_3\chi\\ \chi
\end{array}
\right)e^{i(Q_+z+Et)}
\qquad\mbox{and}\qquad \Psi_-=\left(
\begin{array}{c}
\chi\\-je^{i\phi}\sigma_3\chi
\end{array}
\right)e^{i(Q_-z+Et)}.
\end{equation}
The solution (\ref{Psi_pm})  describes
free particles whose constant spinors depend neither on the energy nor on the potential, and hence cannot recover the 
complex result for $|W_0|=0$. This indicates that the quaternionic Dirac equation may be fundamentally 
different from the complex Dirac equation, and that the complex limit only makes sense in specific cases. Using (\ref{Psi_pm}),
wave-functions composed of particles in both directions for each momentum are
\begin{equation}
\Phi_\pm=\mathcal{A}\,\Psi_\pm(Q_\pm)+\mathcal{B}\,\Psi_\pm(-Q_\pm)
\end{equation} 
where $\mathcal{A}$ and $\mathcal{B}$ are complex constants. Imposing the non-relativistic boundary conditions $\Phi_\pm(z=0)=\Phi_\pm(z=L)=0$, we obtain the quantized momenta
\begin{equation}\label{NR_quant}
 Q_\pm^{(n)}=\frac{n\pi}{L}.
\end{equation}
This result does not necessarily imply that $|W_0|=0$, because $\Phi_+$ and $\Phi_-$ were calculated independently, and we do not
need to impose $Q_+=Q_-$. The relation (\ref{NR_quant}) in fact defines a constraint among the mass, the energy and the 
quaternionic potential parametrized by $|W_0|$ given by
\begin{equation}\label{NR_energy}
E^{(n)2}_\pm=\big(Q_\pm^{(n)}\mp|W_0|\big)^2+m^2.
\end{equation}
(\ref{NR_energy}) may be understood as a relativistic conservation relation for the system, so that 
\begin{equation}
\mathcal{Q}_\pm=Q_\pm^{(n)}\mp|W_0|
\end{equation}
defines an effective momentum. This result squares with the previous result where the velocity of the particle is influenced 
by the quaternionic potential \cite{DeLeo:2013xfa}, so that $|W_0|$ increases $\mathcal{Q_-}$ and decreases $\mathcal{Q_+}$. On the
other hand, it is totally unexpected because we are in fact in the non-relativistic regime.

The effective momentum $\mathcal{Q}_\pm$ has all the properties of the usual relativistic momentum, including a non-relativistic
limit. On the other hand, correspondence between relativistic and non-relativistic quantities is still required, because  better 
comprehension of the quaternion Schr\"odinger equation for confining potentials is needed. Although some progress had already
been made \cite{Nishi:2005qbs,deLeo:2002qpr}, more results are still needed in order to determine the cases in which the quaternionic
quantum results may recover the complex results.

\subsection{The $Q_-$ solution}
We consider the so called ``bag model'' \cite{Chodos:1974je,Chodos:1974pn,Thomas:1982kv}, which has already been used \cite{Fiolhais:1998diw} 
in studying the complex Dirac equation. Defining a space-dependent mass function $\mu(z)$, we get
\begin{equation}\label{M_pot}
\mu(z)=\left\{\;
 m\mbox{ for } 0<z<L\; \mbox{ and }\;M\mbox{ otherwise }\right\}.
\end{equation}
In the limit $M\to\infty$, we have that $Q_\pm^2<0$, and the wave-function becomes zero outside the well.
Inside the square well, where $z\in(0,\,L)$, we have the time independent wave-function
\begin{equation}\label{Q_m_ansatz}
\Psi_-=\mathcal{A}\left[
\begin{array}{c}
\big(1-j\,W_0\,M_-\big)\chi\\\big(A_--j\,W_0\,N_-\big)\sigma_3\chi
\end{array}
\right]e^{iQ_{-}z}+
\mathcal{B}\left[
\begin{array}{c}
\big(1-j\,W_0\,M_-\big)\chi\\\big(-A_-+j\,W_0\,N_-\big)\sigma_3\chi
\end{array}
\right]e^{-iQ_{-}z}
\end{equation}
$\mathcal{A}$ and $\mathcal{B}$ are complex constants. The second term corresponds to a particle moving from the right
to the left, where the momentum is negative.
Taking $Q_-\to-Q_-$ in $\Psi_-$ from (\ref{Psi_pm}), we get $A_-\to-A_-$, $M_-\to M_-$ and $N_-\to-N_-$, so that the second 
term of (\ref{Q_m_ansatz}) is obtained. The model has boundary conditions so that
\begin{equation}\label{bag_boundary}
\Psi=\left\{\;
 \beta\alpha\Psi i\;\mbox{ at }\; z=0\; \mbox{ and }\; -\beta\alpha\Psi i\;\mbox{ at }\; z=L\right\}.
\end{equation}
These boundary conditions are such that the probability current flux is zero outside the well \cite{Fiolhais:1998diw}. We might set 
 $V_0=0$ and thus obtain a pure quaternion potential inside the square well, but we can perform the calculations for a general 
potential and take the limit of a pure quaternionic potential only at the end. Furthermore, we set the spin up solution using $\chi=(1,\,0)^t$,
so that $\sigma_3\chi=\chi$. Thus, from the boundary condition at $z=0$ we get
\begin{equation}
\left[
\begin{array}{c}
(\mathcal{A}+\mathcal{B})(1-jW_0M_-)\chi\\(\mathcal{A}-\mathcal{B})(A_--jW_0N_-)\chi
\end{array}
\right]=
\left[
\begin{array}{c}
(\mathcal{A}-\mathcal{B})(A_--jW_0N_-)\chi\\-(\mathcal{A}+\mathcal{B})(1-jW_0M_-)\chi
\end{array}
\right]i.
\end{equation}
Hence, 
\begin{equation}\label{AB}
\frac{\mathcal{B}}{\mathcal{A}}=\frac{iA_--1}{iA_-+1}=\frac{i\,\overline N_-+\overline M_-}{i\,\overline N_--\overline M_-},
\qquad\mbox{and consequently,}\qquad A_-=-\frac{ \overline N_-}{\overline M_-}.
\end{equation}
This result is absolutely general. However, we must consider the reality of the momentum of the particles.
Real momenta determine that the particle belongs either to the diffusion energy zone or to the Klein energy zone. A pure imaginary
momentum determines that the particle belongs to the evanescent energy zone. However, for the bag model we are considering, 
$M\to\infty$, then the wave-function is zero outside the square well and there is no physical mode on the evanescent zone. 
Real momenta implies that $A_-$ is real from (\ref{AMN}), and then we conclude that
\begin{equation}\label{teta_m}
\frac{\mathcal{B}}{\mathcal{A}}=e^{i\theta}\qquad\mbox{and}\qquad  \tan\theta=\frac{2A_-}{A^2_--1}.
\end{equation}
Where $\theta$ is a phase difference between $\mathcal{A}$ and $\mathcal{B}$. Thus time-independent wave-function is
\begin{equation}\label{Psi_m0}
 \Psi_-=\mathcal{N}_-\left[
\begin{array}{c}
\cos\left(Q_-z-\frac{\theta}{2}\right)-j\, W_0\, M_-\cos\left(Q_-z+\frac{\theta}{2}\right)\\
\left[\sin\left(Q_-z-\frac{\theta}{2}\right)+j\,W_0\,M_-\sin\left(Q_-z+\frac{\theta}{2}\right)\right]iA_-
\end{array}
\right],
\end{equation}
where $\mathcal{N}_-$ is a normalization constant. The boundary condition at $z=0$ for (\ref{Psi_m0}) gives that $\cot\frac{\theta}{2}=A_-$,
in accordance with (\ref{teta_m}). At $z=L$, the boundary condition applied on (\ref{Psi_m0}) gives
\begin{equation}\label{quant_condition}
 \cot\left(Q_-L-\frac{\theta}{2}\right)=-\cot\left(Q_-L+\frac{\theta}{2}\right)=A_-.
\end{equation}
Consequently, the quantization obtained is
\begin{equation}\label{qm_momentum}
 \sin 2Q_-L=0\qquad\mbox{so that}\qquad Q^{(n)}_-=\frac{n\pi}{2L}\qquad\mbox{for}\qquad n\in\mathbb{N}.
\end{equation}
Which obeys the relativistic constraint obtained on the non-relativistic case
\begin{equation}\label{Q_m_energy}
E^{(n)2}_\pm=\big(Q_\pm^{(n)}\mp|W_0|\big)^2+m^2.
\end{equation}
This result squares with the previous result \cite{DeLeo:2013xfa}, where the velocity of the particle is influenced 
by the quaternionic potential, so that $|W_0|$ increases $\mathcal{Q_-}$ and decreases $\mathcal{Q_+}$. 
Additionally, it recovers the non-relativitic limit (\ref{NR_energy}), although the quantized momenta do not match.

In fact, the difference between the quantum results is not important, because the boundary conditions are different in each case,
and so some difference would be expected. However, the result is astonishing  because it differs significantly from the complex 
case. In (\ref{quant_condition})  there are two conditions involved. The condition generated in the complex part of the
wave-function has
$\theta/2$  subtracted in the argument of the trigonometric function and  the condition generated at the quaternion part of the 
wave-function has  $\theta/2$ added to the trigonometric argument. 

The additional quaternionic condition makes the quaternionic result very different from the complex result.
In the complex solution the condition (\ref{quant_condition}) is a transcendental equation, and the quantization is obtained 
numerically \cite{Fiolhais:1998diw}. The quantization condition in the complex case is similar to the non-relativistic {\it finite} 
square well. On the other 
hand, in the quaternionic case, quantization is exact and given by a positive non-zero integer, similar to the  non-relativistic 
{\it infinite} square well, although the difference between the quantized momenta in the quaternion case is half the difference in the
energy on the non-relativistic square well.

From this result it may be interpreted  that the principal difference between the quaternionic calculation and the 
complex calculation is related to its degrees of freedom. At the same time that the quaternion case has more terms to accommodate
the probability density, it may generate more conditions and then constrain the system to a tighter condition compared to the 
complex case. We can speculate that the quaternionic potential may influence the result not
 necessarily as a physical field that can be varied to generate a physical effect, but it can alter the mathematical framework
in a way that the quaternionic effect is not intended to correct the complex case, but to change the phenomenon that can be described.

\subsection{The $Q_+$ solution}
We repeat the calculation for the Dirac solution with momentum $Q_+$, and the results are quite similar.
The ansatz of the time independent wave-function is 
\begin{equation}
\Psi_+=\mathcal{C}\left[
\begin{array}{c}
\big(A_+-j\,\overline W_0\,N_+\big)\sigma_3\chi\\\big(1-j\,\overline W_0\,M_+\big)\chi
\end{array}
\right]e^{iQ_{+}z}+
\mathcal{D}\left[
\begin{array}{c}
\big(-A_++j\,\overline W_0\,N_+\big)\sigma_3\chi\\\big(1-j\,\overline W_0\,M_+\big)\chi
\end{array}
\right]e^{-iQ_{+}z},
\end{equation}
with $\mathcal{C}$ and $\mathcal{D}$ complex constants. At, $z=0$ the boundary condition (\ref{bag_boundary}) on a spin up wave-function gives
\begin{equation}
\left[
\begin{array}{c}
(\mathcal{C}-\mathcal{D})(A_+-j\overline W_0N_+)\chi\\(\mathcal{C}+\mathcal{D})(1-j\overline W_0M_+)\chi
\end{array}
\right]=
\left[
\begin{array}{c}
(\mathcal{C}+\mathcal{D})(1-j\overline W_0M_+)\chi\\(\mathcal{D}-\mathcal{C})(A_+-j\overline W_0N_+)\chi
\end{array}
\right]i.
\end{equation}
\begin{equation}\label{CD}
\frac{\mathcal{D}}{\mathcal{C}}=\frac{iA_++1}{iA_+-1}=\frac{i\,\overline N_+-\overline M_+}{i\,\overline N_++\overline M_+}.
\qquad\mbox{Consequently,}\qquad A_+=-\frac{ \overline N_+}{\overline M_+},\qquad
\frac{\mathcal{D}}{\mathcal{C}}=e^{i\vartheta}\qquad\mbox{and}\qquad  \tan\vartheta=-\frac{2A_+}{A^2_+-1}.
\end{equation}
Where $\vartheta$ is the phase difference between $\mathcal{C}$ and $\mathcal{D}$. Thus time-independent wave-function is
\begin{equation}\label{Psi_p0}
 \Psi_+=\mathcal{N}_+\left[
\begin{array}{c}
iA_+\left[\sin\left(Q_+z-\frac{\vartheta}{2}\right)-j\, \overline W_0\, M_+\sin\left(Q_+z+\frac{\vartheta}{2}\right)\right]\\
\cos\left(Q_+z-\frac{\vartheta}{2}\right)-j\,\overline W_0\,M_+\cos\left(Q_+z+\frac{\vartheta}{2}\right)
\end{array}
\right],
\end{equation}
where $\mathcal{N}_+$ is a normalization constant. The boundary condition at $z=0$ for (\ref{Psi_p0}) gives that $\cot\frac{\theta}{2}=-A_+$,
in accordance with (\ref{CD}). At $z=L$, the boundary condition applied to (\ref{Psi_p0}) gives
\begin{equation}
 -\cot\left(Q_+L-\frac{\theta}{2}\right)=\cot\left(Q_+L+\frac{\theta}{2}\right)=A_+.
\end{equation}
Consequently, the quantization is obtained as
\begin{equation}
 \sin 2Q_+L=0\qquad\mbox{and consequently}\qquad Q_+=\frac{n\pi}{2L}\qquad\mbox{for}\qquad n\in\mathbb{N}.
\end{equation}
This result is absolutely similar to the $Q_-$ case, the difference is that here there is only the diffusion energy zone.

\section{conclusion\label{S4}}
In this article we studied the quaternionic relativistic particle inside an infinite square well. The results show that the quantized
momenta have a quantization condition similar to the energy quantization condition for the {\it infinite} square well. This is very
different from the quantization of the complex Dirac square well, whose quantization of momentum is similar to the energy quantization
of the energy of the {\it finite} square well.

We explain this difference as being due to the greater number of degrees of freedom and of constraints in the quaternionic case.
The result discloses an interesting feature of quaternionic quantum mechanics. At the same time it has more degrees of freedom than 
complex quantum mechanics to accommodate at the
probability density, the quaternionic wave-function duplicates the number of constraints. In the studied case, the new constraints
come from the boundary conditions. The results permit us to state that quaternion quantum mechanics may be useful as a framework 
that permit us to study a physical
system not as a correction of the complex case, but rather as a way of describing different phenomena. This can be concluded from the fact
that the quaternion and the complex cases have quite different solutions for similar problems, and the complex solution cannot
be recovered by simply setting a quaternionic parameter to zero.

The results raises the question about which kind of generalization is promoted by quaternion quantum mechanics. One can consider
the quaternionic part of the wave-function as physically significant or as an additional degree of freedom which may generate a 
mathematical device to describe different phenomena which cannot be described by a complex wave-function. This question must be
addressed using other constrained quantum systems,  both relativistic and non-relativistic.  The finite relativistic potential well
seems probably  the more obvious direction for a future research.

\section*{ACKNOWLEDGEMENTS}
Sergio Giardino receives a financial grant from the CNPq for his research, and is grateful for the hospitality of Professor Paulo 
Vargas Moniz and the Centre for Mathematics and Applications at University of Beira Interior. 
%
%
%
%

\bibliographystyle{unsrt} 
\bibliography{bib_fosso}

\end{document}